\documentclass[floatfix, preprint, onecolumn, prd,showpacs, showkeys, preprintnumbers, nofootinbib, superscriptaddress]{revtex4-1}
\usepackage{times}
\usepackage{amssymb,amsbsy,amsmath,amsfonts}
\usepackage{graphicx}
\usepackage{float}
\usepackage{color}
\usepackage{morefloats}
\usepackage{rotating}
\usepackage{srcltx}
\usepackage{slashed}
\usepackage{multirow}
\usepackage{verbatim}
\usepackage{hyperref}
\usepackage{tabularx}

\begin{document}

\title{Strangeness $S=-1$ hyperon-nucleon interactions: chiral effective field theory vs. lattice QCD}

\author{Jing Song}
\affiliation{School of Physics and Nuclear Energy Engineering and International Research Center for Nuclei and Particles
in the Cosmos, Beihang University, Beijing 100191, China}

\author{Kai-Wen Li}
\affiliation{School of Physics and Nuclear Energy Engineering and International Research Center for Nuclei and Particles
in the Cosmos, Beihang University, Beijing 100191, China}
\affiliation{Yukawa Institute for Theoretical Physics, Kyoto University, Kyoto 606-8502, Japan}

\author{Li-Sheng Geng}
\email[E-mail me at: ]{lisheng.geng@buaa.edu.cn}
\affiliation{School of Physics and Nuclear Energy Engineering and International Research Center for Nuclei and Particles
in the Cosmos, Beihang University, Beijing 100191, China}
\affiliation{Beijing Key Laboratory of Advanced Nuclear Materials and Physics, Beihang University, Beijing 100191, China}
\affiliation{Beijing Advanced Innovation Center for Big Date-based Precision Medicine, Beihang University, Beijing, 100191,China}

\begin{abstract}
Hyperon-nucleon interactions serve as basic inputs to studies of hypernuclear physics and dense (neutron) stars. Unfortunately, a precise understanding of these important quantities have lagged far behind that of the nucleon-nucleon interaction due to lack of high precision experimental data. Historically, hyperon-nucleon interactions are either formulated in quark models or meson exchange models. In recent years, lattice QCD simulations and chiral effective field theory approaches start to offer new insights from first principles. In the present work, we contrast the state of art lattice QCD simulations with the latest chiral hyperon-nucleon forces and show that the leading order relativistic chiral results can already describe the lattice QCD data reasonably well. Given the fact that the lattice QCD simulations are performed with pion masses ranging from the (almost) physical point to 700 MeV, such studies provide a highly non-trivial check on both the chiral effective field theory approaches as well as lattice QCD simulations. Nevertheless more precise lattice QCD simulations are eagerly needed to refine our understanding of hyperon-nucleon interactions.
\end{abstract}

\pacs{}
\keywords{}

\date{\today}

\maketitle
\section{Introduction}
Baryon-baryon (nucleon-nucleon, hyperon-nucleon, hyperon-hyperon) interactions play fundamental roles in microscopic studies of nuclear physics and astro-nuclear physics~\cite{Doi:2017niz, Gal:2016boi}. Many outstanding issues in these fields, such as the existence of the H-dibaryon~\cite{Jaffe:1976yi}, the large charge symmetry breaking in $^4_\Lambda$H and $^4_\Lambda$He~\cite{Yamamoto:2015avw} and the \textit{hyperon puzzle} for neutron stars~\cite{Massot:2012pf,Schulze:2011zza,Hu:2013tma,Miyatsu:2013hea}, are due at least partly to the difficulty in calculating them from first principles and partly due to the scarce experimental data on hyperon-nucleon ($YN$) scattering in the low energy region. Conventionally, $YN$ interactions are constructed in either quark models or meson-exchange models. Their connection to the underlying theory of the strong interaction, QCD, however, is not very transparent. In the past two decades, studies based on lattice QCD and chiral effective field theory (ChEFT) have made remarkable progress and provided new insights on these rather old but extremely important quantities.

As a brute force numerical solution of QCD, lattice QCD can in principle solve low-energy strong interaction physics from first principles. In practice, there are still many issues to be solved. For instance, until very recently, lattice QCD simulations are usually performed with larger than physical light quark (i.e., pion) masses. On the other hand, for simulations at the (almost) physical point, statistical uncertainties become increasingly large~\cite{Doi:2017zov}.  ChEFT, being also model independent, relies on either experimental data or lattice QCD simulations to better constrain the relevant low energy constants (LECs). Lack of such information can result in considerable uncertainties in the constructed potentials.

In the present work, we contrast the latest chiral $YN$ interactions with the state of art lattice QCD simulations. In Sec. II, we briefly explain the general features of the relativistic chiral $YN$ interactions. Then we contrast them with the three lattice QCD simulations of Refs.~\cite{Beane:2006gf,Miyamoto:2016hqo,Nemura:2017vjc} and we show that the agreement seems to be reasonable, though some discrepancies remain. We stress that such agreements provide a highly nontrivial check on the ChEFT approaches as well as lattice QCD simulations. Then we end with a short summary.

\section{HYPERON-NUCLEON INTERACTIONS IN CHIRAL EFT}

Up to now, three ChEFT approaches have been explored to study the $YN$ system. The PecsGroningen group used the KSW (Kaplan-Savage-Wise) approach to study the strangeness $S = -1$ $YN$ interactions up to next-to-leading order (NLO) ~\cite{Korpa:2001au}. However, the KSW approach is known to suffer from slow convergence in the nucleon-nucleon $NN$ sector~\cite{Fleming:1999bs}. The Bonn-J\"{u}lich group used the heavy baryon (HB) approach, which has achieved remarkable successes in the $NN$ sector ~~\cite{Bedaque:2002mn,Epelbaum:2008ga,Machleidt:2011zz}, to study the strangeness $S=-1$ and $S=-2$ systems up to NLO ~\cite{Polinder:2006zh,Haidenbauer:2007ra,Haidenbauer:2013oca,Polinder:2007mp,Haidenbauer:2015zqb}, and the $S=-3$ and $S=-4$ systems at leading order (LO)~\cite{Haidenbauer:2009qn}. Recently, the Beijing-Chengdu group has proposed a covariant power counting scheme to study the $NN$~\cite{Ren:2016jna} and strangeness $S=-1$ $YN$ systems~\cite{Li:2016mln}. The main features of this new approach are that the small component of the baryon spinor is retained and  a relativistic scattering equation (i.e., the Kadyshevsky equation) is solved to iterate the baryon-baryon potentials. It is shown that already at LO, this new formalism can describe the $NN$ phase shifts and $YN$ cross sections fairly well, comparable with the NLO HB approach.

Here we would like to briefly summarize the essential ingredients of the relativistic ChEFT approach for baryon-baryon interactions. Details can be found in Refs.~\cite{Ren:2016jna,Li:2016mln}. The LO baryon-baryon potentials consist of non-derivative four-baryon contact terms (CT) and one-pseudoscalar-meson exchange terms (OPME). The corresponding Lagrangian for the CT reads,
 \begin{align}\label{CT}
  \mathcal{L}_{\textrm{CT}} = \sum_{i=1}^5\left[\frac{\tilde C_i^1}{2}~\textrm{tr}\left(\bar B_1 \bar B_2 (\Gamma_i B)_2 (\Gamma_i B)_1\right)
  + \frac{\tilde C_i^2}{2}~\textrm{tr}\left(\bar B_1 (\Gamma_i B)_1 \bar B_2 (\Gamma_i B)_2\right)
  + \frac{\tilde C_i^3}{2}~\textrm{tr}\left(\bar B_1 (\Gamma_i B)_1\right)\textrm{tr}\left( \bar B_2 (\Gamma_i B)_2\right)\right],
\end{align}
where $\mathrm{tr}$ indicates the trace in flavor space ($u$, $d$, and $s$).  $\Gamma_i$ are the elements of the Clifford algebra. $B$ is a $3\times3$ traceless matrix collecting the ground-state octet baryons. The OPME Lagrangian is
\begin{align}\label{LMB1}
  &\mathcal{L}_{MB}^{(1)} =
  \mathrm{tr}\Bigg( \bar B \big(i\gamma_\mu D^\mu - M_B \big)B -\frac{D}{2} \bar B \gamma^\mu\gamma_5\{u_\mu,B\}
  -  \frac{F}{2}\bar{B} \gamma^\mu\gamma_5 [u_\mu,B]\Bigg)\, ,
\end{align}
where $D^\mu B = \partial_\mu B+[\Gamma_\mu,B] $ and $D$ and $F$ are the axial vector couplings. $\Gamma_\mu$  and $u_\mu$ are the vector and axial vector combinations of the pseudoscalar-meson fields and their derivatives,
\[
  \Gamma_\mu = \frac{1}{2}\left(u^\dag\partial_\mu u + u\partial_\mu u^\dag \right), \quad u_\mu=i(u^\dagger \partial_\mu u-u\partial_\mu u^\dagger)\, ,
\]
where $u^2= U = \exp\left(i\frac{\sqrt{2}\phi}{f_0}\right)$, and the traceless matrix $\phi$ collects the pseudoscalar-meson fields. In the derivation of the baryon-baryon potentials, the complete baryon spinor has been used,
\begin{equation}\label{ub}
  u_B(\mbox{\boldmath $p$}, s)= N_p
  \left(
  \begin{array}{c}
    1 \\
    \frac{\mbox{\boldmath $\sigma$}\cdot \mbox{\boldmath $p$}}{E_p+M_B}
  \end{array}\right)
  \chi_s,
  ~~~~
  N_p=\sqrt{\frac{E_p+M_B}{2M_B}},
\end{equation}
where $E_p=\sqrt{\mbox{\boldmath $p$}^2+M_B^2}$, while a non-relativistic reduction of $u_B$ is employed in the HB and KSW approaches. In addition, the coupled-channel Kadyshevsky equation is solved to take into account the non-perturbative nature of the baryon-baryon interactions,
\begin{align}\label{SEK}
  & T_{\rho\rho'}^{\nu\nu',J}(\mbox{\boldmath $p$}',\mbox{\boldmath $p$};\sqrt{s})
  =
   V_{\rho\rho'}^{\nu\nu',J}(\mbox{\boldmath $p$}',\mbox{\boldmath $p$})
   +
  \sum_{\rho'',\nu''}\int_0^\infty \frac{dp''p''^2}{(2\pi)^3} \frac{M_{B_{1,\nu''}}M_{B_{2,\nu''}}~ V_{\rho\rho''}^{\nu\nu'',J}(\mbox{\boldmath $p$}',\mbox{\boldmath $p$}'')~
   T_{\rho''\rho'}^{\nu''\nu',J}(\mbox{\boldmath $p$}'',\mbox{\boldmath $p$};\sqrt{s})}{E_{1,\nu''}E_{2,\nu''}
  \left(\sqrt{s}-E_{1,\nu''}-E_{2,\nu''}+i\epsilon\right)},
\end{align}
where $\sqrt{s}$ is the total energy of the two-baryon system in the center-of-mass frame and $E_{n,\nu''}=\sqrt{\mbox{\boldmath $p$}''+M_{B_{n,\nu''}}}$, $(n=1,2)$. The labels $\nu,\nu',\nu''$ denote the particle channels, and $\rho,\rho',\rho''$ denote the partial waves. In numerical evaluations, the potentials in the scattering equation are regularized with an exponential form factor,
\begin{equation}\label{EF}
  f_{\Lambda_F}(\mbox{\boldmath $p$},\mbox{\boldmath $p$}') = \exp \left[-\left(\frac{\mbox{\boldmath $p$}}{\Lambda_F}\right)^{4}-\left(\frac{\mbox{\boldmath $p$}'}{\Lambda_F}\right)^{4}\right] \, .
\end{equation}

A rather good description of the $\Lambda N$ and $\Sigma N$ cross sections was achieved in Ref.~\cite{Li:2016mln} with a cutoff $\Lambda_F$ = 600 MeV. In the present work, we contrast the results of Ref.~\cite{Li:2016mln} with the three lattice QCD simulations of Refs.~\cite{Beane:2006gf,Miyamoto:2016hqo,Nemura:2017vjc}. These will provide a highly non-trivial check on the potentials obtained there.

\section{HYPERON-NUCLEON INTERACTIONS IN LATTICE QCD SIMULATIONS}

Lattice QCD simulations of the $NN$ and $YN$ interactions are mainly being pursued by two collaborations, i.e., the HAL QCD collaboration and the NPLQCD collaboration. The HAL QCD collaboration developed the so-called HAL QCD method to extract potentials from lattice QCD simulations, while the NPLQCD collaboration relies on the L\"{u}scher method to obtain phase shifts or binding energies directly. At present, there are some heated discussions about the pros and drawbacks of both methods, see, e.g., Ref.~\cite{Aoki:2017yru}. In the $S=-1$ sector, there are mainly three lattice QCD simulations, i.e., Refs.~\cite{Beane:2006gf,Miyamoto:2016hqo,Nemura:2017vjc}, which will be the focus of the present work,

In Ref.~\cite{Beane:2006gf}, the NPLQCD collaboration performed fully-dynamical lattice QCD simulations of $\Lambda n$ and $\Sigma^-n$ scattering lengths and phase shifts in the $^1S_0$ and $^3S_1$ channel at three pion masses and six center-of-mass momenta. The simulations were carried out with
the lattice configurations detailed in Ref.~\cite{WalkerLoud:2008bp}. The corresponding  pseudoscalar and baryon masses  (in units of MeV) \cite{WalkerLoud:2008bp} are collected in Table ~\ref{2}.


\begin{table}[htpb]
   \caption{ Masses of the pseudoscalar mesons and octet baryons of the LHPC~\cite{WalkerLoud:2008bp}. The first error is the statistical uncertainty and the second is determined by the lattice spacing. Those denoted by stars were used
   by the NPLQCD Collaboration~\cite{Beane:2006gf}.}\label{2}
\centering
\newcommand{\tabincell}[2]{\begin{tabular}{@{}#1@{}}#2\end{tabular}}
\begin{tabular}{cc|ccccc}
\hline
\hline
$M_\pi$(MeV) & $M_K$(MeV) & $m_N$(MeV) & $m_\Lambda$(MeV) & $m_\Sigma$(MeV) & $m_\Xi$(MeV) \\
 \hline

292.9 & 585.6 & 1098.9(8.0)(22.0) & 1240.5(4.8)(24.8) & 1321.6(6.4)(26.4) & 1412.2(3.2)(28.2) \\

$355.9^*$ & 602.9 & 1157.8(6.4)(23.1) & 1280.2(4.8)(25.6) & 1350.2(4.8)(27.0) & 1432.9(3.2)(28.6)\\

$495.1^*$ & 645.2 & 1288.2(6.4)(25.8) & 1369.3(4.8)(27.4) & 1409.1(6.4)(28.2) & 1469.5(4.8)(29.4)\\

$596.7^*$ & 685.6 & 1394.8(6.4)(27.9) & 1440.9(8.0)(28.8) & 1463.1(9.5)(29.2) & 1504.5(8.0)(30.1)\\

687.7 & 728.1 & 1502.9(11.1)(30.0) & 1528.3(9.5)(30.6) & 1536.3(9.5)(30.7) & 1557.0(9.5)(31.1)\\
\hline
\hline
\end{tabular}

\end{table}

\begin{table}[htpb]
   \caption{ Masses of the pseudoscalar mesons and octet baryons of the PACS-CS Collaboration~\cite{Aoki:2008sm}. The first error is the statistical uncertainty and the second is determined by the lattice spacing. Those denoted by stars were
   used by the HAL QCD Collaboration~\cite{Miyamoto:2016hqo}.}\label{3}
\centering
\newcommand{\tabincell}[2]{\begin{tabular}{@{}#1@{}}#2\end{tabular}}
\begin{tabular}{cc|ccccc}
\hline
\hline
$M_\pi$(MeV) & $M_K$(MeV) & $m_N$(MeV) & $m_\Lambda$(MeV) & $m_\Sigma$(MeV) & $m_\Xi$(MeV) \\
 \hline

155.8 & 553.7 & 932.1(78.3)(14.4) & 1139.9(20.7)(17.6) & 1218.4(21.5)(18.8) & 1393.3(6.7)(21.5) \\

295.7 & 593.5 & 1093.1(18.9)(16.9) & 1253.8(14.1)(19.4) & 1314.8(15.4)(20.3) & 1447.7(10.0)(22.3)\\

384.4 & 581.4 & 1159.7(15.4)(17.9) & 1274.1(9.1)(19.7) & 1316.5(10.4)(20.3) & 1408.3(7.0)(21.7)\\

411.2 & 635.0 & 1214.7(11.5)(18.7) & 1350.4(7.8)(20.8) & 1400.2(8.5)(21.6) & 1503.1(6.5)(23.2)\\

$569.7^*$ & 713.2 & 1411.1(12.2)(21.8) & 1503.8(9.8)(23.2) & 1531.2(11.1)(23.6) & 1609.5(9.4)(24.8)\\

$701.4^*$ &789.0 & 1583.0(4.8)(24.4) & 1643.9(5.0)(25.4) & 1654.5(4.4)(25.5) & 1709.6(5.4)(26.4)\\
\hline
\hline
\end{tabular}

\end{table}

In Ref.~\cite{Miyamoto:2016hqo}, the HAL QCD collaboration studied the $\Lambda N$ phase shifts and scattering lengths in the $^1S_0$ channel at two pion masses, $m_\pi=700(1)$ MeV and $m_\pi=570(1)$ MeV. They also obtained the $\Lambda N$ phase shifts as a function of the center-of-mass energy in the range of $0\leq \textrm{E}_{\textrm{cm}} \leq 200$ MeV.  The details of the  lattice QCD configurations can be found in Ref.~\cite{Aoki:2008sm}. In Table~\ref{3}, we list the corresponding pseudoscalar and baryon masses.

%

In Ref.~\cite{Nemura:2017vjc} the HAL QCD collaboration studied the $\Sigma N (I = 3/2)$ scattering phase shifts in the $^3S_1$ - $^3D_1$ and $^1S_0$ channel for the first time with an almost physical pion mass ($m_\pi \approx$ 146 MeV), though with rather large statistical uncertainties.

\section{RESULTS AND DISCUSSION}

In Refs.~\cite{Beane:2006gf,Miyamoto:2016hqo}, the lattice QCD simulations are performed with larger than physical light quark masses (i.e., pion masses). Only in Ref.~\cite{Nemura:2017vjc}, simulations are performed with an almost physical pion mass. In ChEFT studies, one normally assumes that the LECs are quark mass independent, therefore all the quark mass dependence comes form that of the masses of the interacting baryons, as well as the exchanged mesons~\footnote{Strictly speaking, this is only true for leading order studies. At higher chiral orders, the quark mass dependence of
the couplings will contribute, see, e.g., Refs.~\cite{Beane:2002vs,Epelbaum:2002gb}.}. Once the dependence is known, one simply performs a chiral extrapolation to extrapolate lattice QCD simulations performed at unphysical light quark masses to the physical point. Such a procedure has turned out be quite successful in the one-baryon sector, see, e.g., the baryon masses ~\cite{Ren:2012aj,Ren:2014vea}. Similar studies have been performed in the baryon-baryon sector ~\cite{Haidenbauer:2017sws}. Nevertheless, as known in the one-baryon sector, even at next-to-next-to-next-to-leading order, the validity of BChPT is somehow limited to up to $m_\pi\sim$ 500 MeV. In principle, it is not clear whether one can apply the LO ChEFT to extrapolate the LQCD simulations of Refs.~\cite{Beane:2006gf,Miyamoto:2016hqo}

In the present work, we fix the pion mass dependence of the kaon(eta) and octet baryons by fitting to the $n_f=2+1$ PACS-CS and LHPC lattice QCD data using the LO and NLO chiral perturbation theory. The kaon mass (in units of MeV) is related to that of the pion via ~\cite{Ren:2012aj}
\begin{equation*}
m_K^2 = 0.291751 + 0.670652 m_\pi^2\quad\mbox{for the PACS-CS configurations,}
\end{equation*}
\begin{equation*}
m_K^2 = 0.301239 + 0.479545 m_\pi^2\quad\mbox{for the LHPC configurations.}
\end{equation*}

The octet baryon masses up to the second order in the chiral expansion read,\\
\begin{equation*}
m_B=m_0+m_B^{(2)}.
\end{equation*}
At $\mathcal{O}(p^2)$ the tree level contribution provides the LO SU(3)-breaking corrections to the chiral limit octet baryon mass\\
\begin{equation*}
m_B^{(2)}=\sum\limits_{\phi=\pi,K}\xi_{B,\phi}^{(a)}M_\phi^2
\end{equation*}
where the coefficients $\xi_{B,\phi}^{(a)}$ can be found in Ref.~\cite{Ren:2012aj}. The LECs can be
determined by a least-squares fit to the baryon masses given in Tables \ref{2} and \ref{3} and are tabulated in Tables~\ref{5} and ~\ref{6}.

\begin{table}[t]
   \caption{Values of the NLO LECs determined by fittiing to the PACS-CS baryon masses.}\label{5}
\newcommand{\tabincell}[2]{\begin{tabular}{@{}#1@{}}#2\end{tabular}}
  \centering
\begin{tabular}{cccc}
\hline
\hline
$m_0[\textrm{MeV}]$& $b_0[\textrm{GeV}^{-1}]$ & $b_D[\textrm{GeV}^{-1}]$ & $b_F[\textrm{GeV}^{-1}]$ \\
 \hline
962.32 & $-0.22809$ & 0.038183 & $-0.15768$ \\
\hline
\hline
   \end{tabular}
\end{table}

\begin{table}[t]
   \caption{Values of the LECs determined by fitting to the LHPC baryon masses.}\label{6}
\newcommand{\tabincell}[2]{\begin{tabular}{@{}#1@{}}#2\end{tabular}}
  \centering
\begin{tabular}{cccc}
\hline
\hline
$m_0[\textrm{MeV}]$& $b_0[\textrm{GeV}^{-1}]$ & $b_D[\textrm{GeV}^{-1}]$ & $b_F[\textrm{GeV}^{-1}]$ \\
 \hline
1026.1 & $-0.20255 $ & 0.054454 & $-0.14462$ \\
\hline
\hline
   \end{tabular}

\end{table}

\begin{figure}[b]
  \centering
\includegraphics[width=\textwidth]{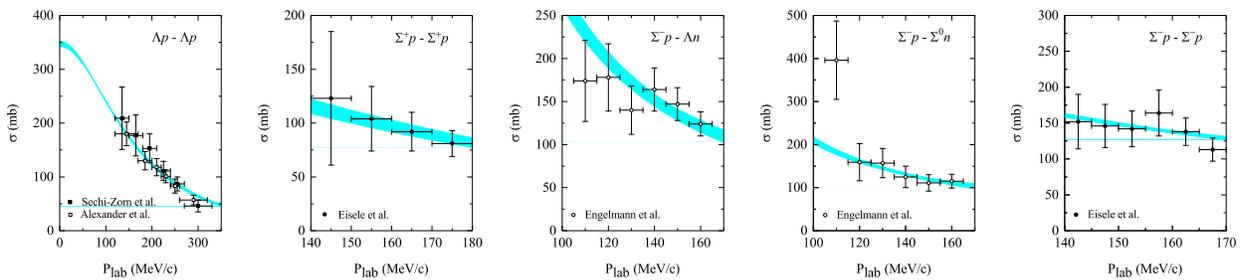}
 \caption{$\Lambda N$ and $\Sigma N$ cross sections as functions of the laboratory momentum. The bands correspond to results obtained with cutoffs ranging from 550 to 850 MeV, while in Ref.~\cite{Li:2016mln} only the best fit with a cutoff of 600 MeV was shown. The experimental data are taken from Sechi-Zorn et al.~\cite{SechiZorn:1969hk}, Alexander et al.~\cite{Alexander:1969cx}, Engelmann et al.~\cite{Engelmann:1966}, and Eisele et al.~\cite{Eisele:1971mk}.}\label{7}
\end{figure}

Before comparing the ChEFT results with the lattice QCD simulations, we point out that in Ref.~\cite{Li:2016mln}, one obtained the best description of the experimental data at a cutoff of 600 MeV and with a $\chi^2$ about 16. Given the number of degrees of freedom is 20, it is clear that with cutoffs ranging from 550 MeV to 850 MeV the $\chi^2/\textrm{d.o.f}$ is always smaller than 1 and as a result they cannot be distinguished from each other from a statistical point of view. This is shown in Fig.~\ref{7}. Indeed, all the theoretical results agree with the experimental data within uncertainties.

\begin{figure}[htpb]
  \centering
\includegraphics[width=\textwidth]{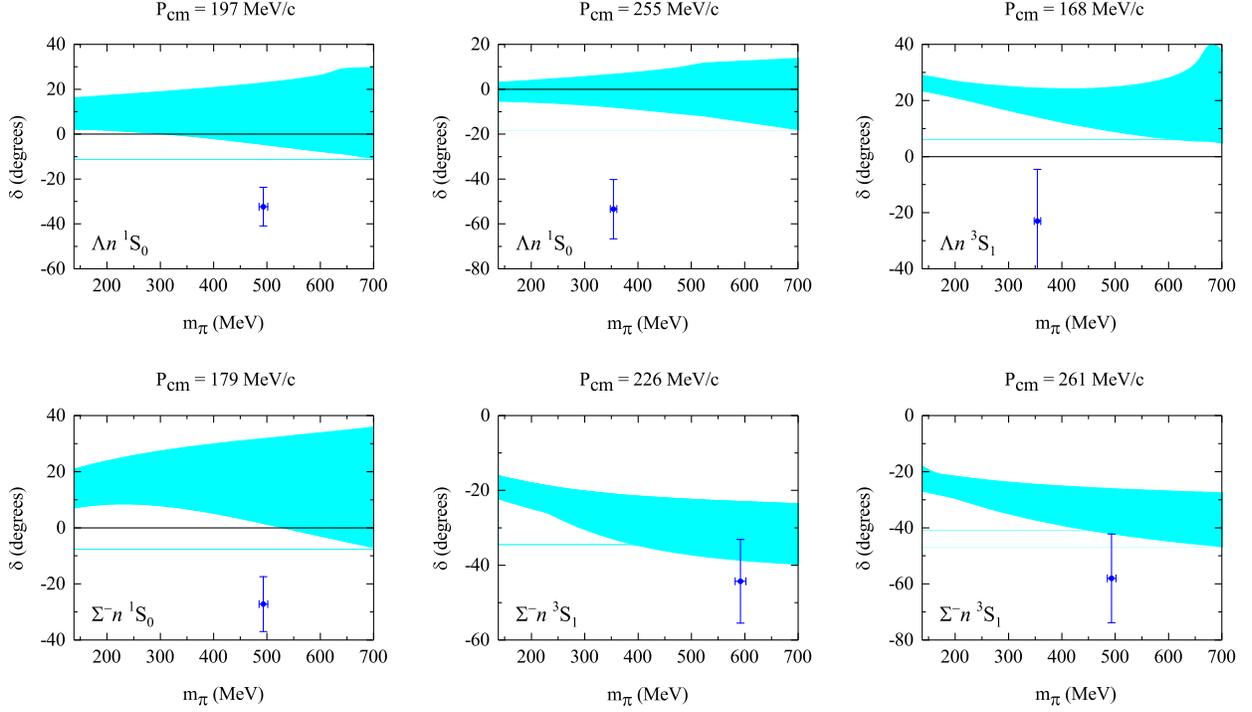}
 \caption{NPLQCD ~\cite{Beane:2006gf}  $\Lambda n$ and $\Sigma^-n$ phase shifts (blue filled circles) as a function of the pion mass in comparison with the ChEFT predictions. The cyan bands correspond to the results obtained with cutoffs ranging from 550 to 850 MeV.}\label{8}
\end{figure}
\begin{figure}[htpb]
\centering
\includegraphics[width=\textwidth]{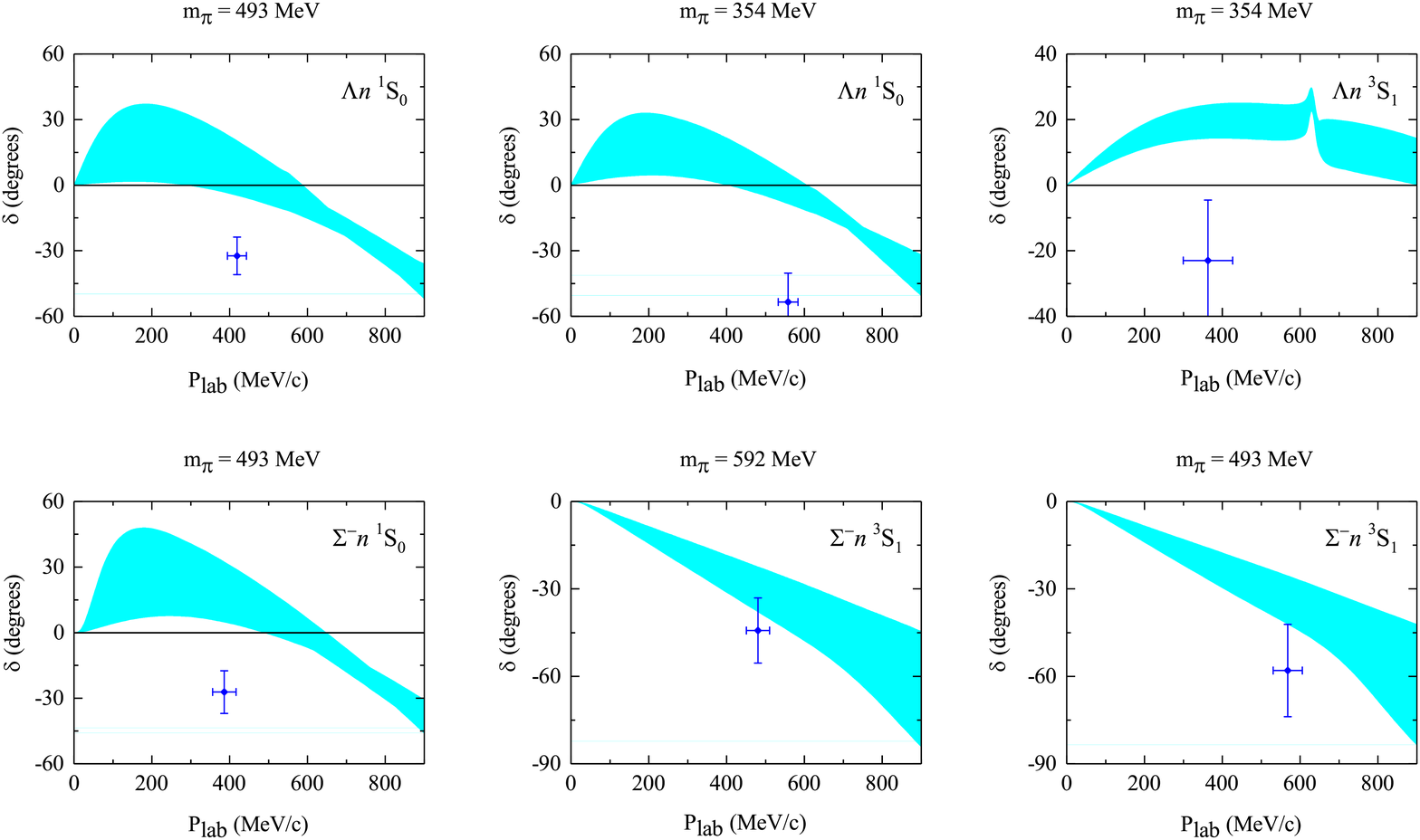}
 \caption{NPLQCD ~\cite{Beane:2006gf} 
$\Lambda n$ and $\Sigma^-n$ phase shifts (blue filled circles)  as a function of the laboratory momentum $P_\textrm{{lab}}$ in comparison with the ChEFT predictions. The bands correspond to the variation of the cutoff $\Lambda_F$ from 550 to 850 MeV.}\label{9}
\end{figure}

In Figs.~\ref{8} and \ref{9}, we compare the NPLQCD  phase shifts obtained at different pion masses and center-of-mass momenta with the ChEFT predictions.   It is clear that except in the two lower-left panels, the ChEFT results do not agree with the lattice QCD data, even taking into account cutoff uncertainties. This may not be a complete surprise because of the following two reasons. First, the lattice QCD simulations are performed not only with larger than physical pion masses (the smallest one being 354 MeV) but also with relatively large laboratory momentum (the smallest being  357 MeV/c). Particularly, in Ref.~\cite{Li:2016mln}, only the scattering data with $P_\textrm{{lab}} \le 300$ MeV/c were fitted. Second, the ChEFT study is only a LO study. One probably needs to go to the NLO to properly describe the NPLQCD data. One caveat in the above comparison is that we used the LO/NLO BChPT to describe the light quark mass dependence of the pseudoscalar and baryon masses, which may not be able to completely describe the lattice QCD data (e.g., the finite volume effects). However, we have checked using the baryon masses provided by the lattice QCD collaborations themselves does not lead to appreciable differences.

\begin{figure}[htpb]
  \centering
\includegraphics[width=\textwidth]{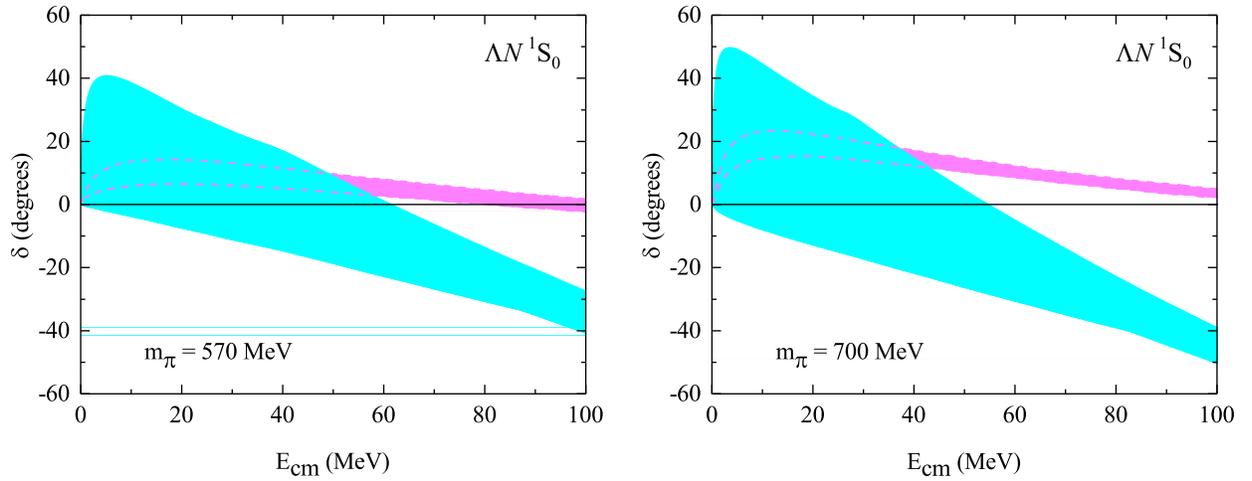}
 \caption{HAL QCD ~\cite{Miyamoto:2016hqo} $\Lambda N$ phase shifts   (magenta bands)  in comparison with the ChEFT predictions. The cyan bands correspond to the results obtained with cutoffs ranging from 550 to 850 MeV.}\label{11}
\end{figure}

In Fig.~\ref{11}, we compare the ChEFT predicted phase shifts with the HAL QCD results of Ref.~\cite{Miyamoto:2016hqo}. For $E_\mathrm{cm}< 40$ MeV, which corresponds to $P_{\textrm{lab}}\le$ 506 MeV/c  ($m_\pi=570$ MeV)  or 527 MeV/c ($m_\pi=700$ MeV), the ChEFT phase shifts agree with the lattice QCD data within uncertainties. For larger $E_\mathrm{cm}$, the ChEFT phase shifts decrease faster the lattice QCD results. Again, this might indicate that one needs to go to higher orders to properly describe the lattice QCD data with large pion masses and large $P_{\textrm{lab}}$.

\begin{figure}[t]
\centering
\includegraphics[width=\textwidth]{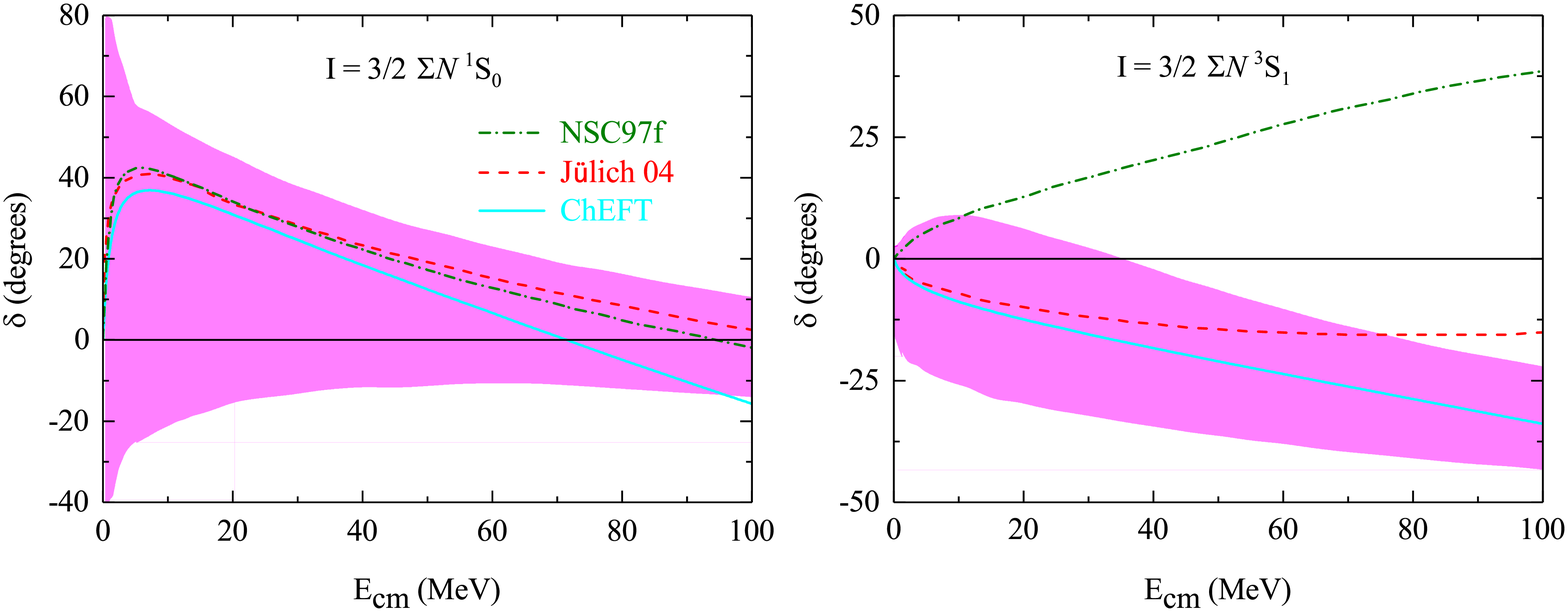}
 \caption{
 HAL QCD data~\cite{Nemura:2017vjc} in comparison with the ChEFT, J\"{u}lich04~\cite{Haidenbauer:2005zh} and NSC97f~\cite{Rijken:1998yy} phase shifts. }\label{10}
\end{figure}

The HAL QCD results with an almost physical pion are compared with the ChEFT results as well as those of the J\"{u}lich04~\cite{Haidenbauer:2005zh} and NSC97f~\cite{Rijken:1998yy}~\footnote{It should be noted that the J\"{u}lich04 and NSC97f results are in fact for $\Sigma^+ p$, not for $\Sigma^- n$. We have checked that the Coulomb effects are small and do not alter the qualitative comparison shown here.} in Fig.~\ref{10}. It is clear that the ChEFT results agree with the lattice QCD data quite well, albeit the uncertainties of the preliminary lattice QCD data are very large.  On the other hand, the J\"{u}lich04 phase shifts also agree with the lattice QCD simulations reasonably well. However, the NSC97f phase shifts only agree with the lattice QCD data  in the $^1S_0$ channel, but not in the $^3S_1$ channel.  Clearly, more refined lattice QCD simulations will provide valuable constraints on the different formulations of the $YN$ interactions.

\section{Conclusion}

We have extrapolated the hyperon-nucleon interactions in the relativistic chiral effective field theory to unphysical pion masses. Using the next-to-leading order (leading order) chiral perturbation theory to describe the light quark mass dependence of the octet baryons (the kaon/eta), we found that the predicted phase shifts at unphysical light quark masses are in reasonable agreement with the lattice QCD simulations, particularly with those of the HAL QCD collaboration at small laboratory momenta. Nevertheless, some discrepancies remain, which call for higher order studies in the chiral effective field theory approaches.

It should be stressed that the present study better be viewed as of exploratory nature. Once more refined lattice QCD studies become available, one may work out the relativistic chiral forces up to the next-to-leading order and study theoretical uncertainties more carefully. Nevertheless, the present study provides a highly non-trivial check on the relativistic chiral hyperon-nucleon interactions.

\section{Acknowledgements}
This work was supported in part by the National Natural Science Foundation of China under Grants No. 11522539 and No. 11735003. K.-W.L. acknowledges financial support from the China Scholarship Council.

\end{document}